\documentclass[pdflatex,sn-mathphys-num]{sn-jnl}

\usepackage{graphicx}%
\usepackage{multirow}%
\usepackage{amsmath,amssymb,amsfonts}%
\usepackage{amsthm}%
\usepackage{mathrsfs}%
\usepackage[title]{appendix}%
\usepackage{xcolor}%
\usepackage{textcomp}%
\usepackage{manyfoot}%
\usepackage{booktabs}%
\usepackage{algorithm}%
\usepackage{algorithmicx}%
\usepackage{algpseudocode}%
\usepackage{listings}%

\theoremstyle{thmstyleone}%
\newtheorem{theorem}{Theorem}
\newtheorem{Observation}[theorem]{Observation}%

\theoremstyle{thmstyletwo}%

\theoremstyle{thmstylethree}%

\begin{document}

\title[Article Title]{A Divide-and-Conquer Engine for Lexicographical Permutations: Accelerating State Evolution via Hybrid Software-Hardware CPU Instructions}


\author*[1]{\fnm{Yusheng} \sur{HU}}  \email{dr.huyusheng@gmail.com}


\affil*[1]{\orgdiv{Independent Researcher}, \orgaddress{\street{Shangdi East Road}, \city{Beijing}, \postcode{100085}, \country{China}}}

\abstract{Traditional lexicographical permutation algorithms, epitomized by \texttt{std::next\_permutation}, are fundamentally bottlenecked by dense control logic and frequent branch mispredictions. This paper introduces LexCHA(Lexicographical Co-designed Hardware Acceleration), a novel hardware-software co-designed architecture that accelerates permutation via native SIMD instructions. Exploiting the inherent fractal isomorphism of permutations, LexCHA decouples global state evolution into a macro software unranking phase and a micro hardware block-construction phase. By utilizing pre-computed deterministic mask streams, LexCHA replaces traditional conditional branching with streaming vector shuffles, achieving a branch-free execution flow. Evaluations demonstrate that LexCHA demonstrates significantly higher throughput than scalar implementations of modern processors, outperforming existing state-of-the-art \texttt{std::next\_permutation} implementations significantly.}



\maketitle

\section{Introduction}

Lexicographical permutation generation is a fundamental primitive in combinatorial computing. However, traditional algorithms, such as the Narayana-Pandita approach, often face challenges when mapped to modern superscalar architectures, which can impact overall system throughput. Their reliance on bidirectional scans and nested loops tends to favor serial execution and frequent conditional branches, which may lead to branch mispredictions and pipeline stalls on deeply pipelined processors.

In the study of permutation generation algorithms, the lexicographic order algorithm is generally considered to be less efficient in terms of both time complexity and implementation performance compared to Heap's algorithm \cite{Heap1963} and the adjacent transposition-based Steinhaus-Johnson-Trotter (SJT) algorithm \cite{Johnson1963}.

Building upon the theoretical framework of factorial-based permutation generation \cite{Arndt2010}, we observe that traditional iterative methods are constrained by sequential execution overhead and branch misprediction. To overcome these limitations, this paper proposes LexCHA (Lexicographical Co-designed Hardware Architecture), a novel architecture that accelerates permutation generation through software-hardware co-design. By decoupling global state transitions into a dual-layered paradigm and utilizing precomputed deterministic execution masks, LexCHA effectively bypasses the throughput limitations of conventional branch-heavy implementations.

At the microarchitectural level, LexCHA implements the core algorithmic logic using native SIMD vector primitives, effectively removing runtime branches and algorithmic redundancy to achieve a streamlined, dataflow-centric execution.

\section{Deterministic topological features of lexicographical order evolution}
This chapter demonstrates the topological structure of the permutation generation algorithm and reveals that its logical complexity can be transformed into branchless data flow computation through fractal isomorphism.

\subsection{Topological deconstruction of permutation space}
Traditional lexicographical order generation algorithms rely heavily on the current permutation state to determine subsequent transformations. This dependency causes both the number of available candidate elements and the length of the remaining search path to fluctuate dynamically as the algorithm progresses. This paper models permutation generation as a discrete state traversal on a symmetric group topological space, proving that the evolution between any adjacent states has a deterministic mapping path, providing a theoretical basis for eliminating branching logic.

\subsubsection{Definition of the Factorial Number System}
Based on the classical approach to permutation encoding introduced by Lehmer~\cite{Lehmer1960}, we define the structure of the factorial array $C$ to establish a bijective mapping between a permutation and its lexicographical rank:

\[
\begin{array}{r@{\quad}l}
	C[0]   &\in \{0, 1, \dots, n-1\} \\
	C[1]   &\in \{0, 1, \dots, n-2\} \\
	\multicolumn{2}{c}{\vdots} \\
	C[n-2] &\in \{0, 1\} \\
	C[n-1] &\in \{0\}
\end{array}
\]

The traversal of the factorial array follows a standard lexicographical increment, evolving monotonically from the initial state:
\[
\mathbf{0} = (0, 0, \dots, 0, 0)
\]
to the terminal state:
\[
\mathbf{M} = (n-1, n-2, \dots, 1, 0)
\]

\begin{table}[htbp]
	\centering
	\caption{Full Lexicographical State Mapping for $n=4$}
	\label{tab:lex_mapping_n4}
	\small
	\begin{tabular}{ccccccccccc}
		\toprule
		\textbf{Rank} & \textbf{Fact.} & \textbf{Perm.} & & \textbf{Rank} & \textbf{Fact.} & \textbf{Perm.} & & \textbf{Rank} & \textbf{Fact.} & \textbf{Perm.} \\
		\midrule
		0  & 0000 & 1234 & & 8  & 1100 & 2314 & & 16 & 2200 & 3412 \\
		1  & 0010 & 1243 & & 9  & 1110 & 2341 & & 17 & 2210 & 3421 \\
		2  & 0100 & 1324 & & 10 & 1200 & 2413 & & 18 & 3000 & 4123 \\
		3  & 0110 & 1342 & & 11 & 1210 & 2431 & & 19 & 3010 & 4132 \\
		4  & 0200 & 1423 & & 12 & 2000 & 3124 & & 20 & 3100 & 4213 \\
		5  & 0210 & 1432 & & 13 & 2010 & 3142 & & 21 & 3110 & 4231 \\
		6  & 1000 & 2134 & & 14 & 2100 & 3214 & & 22 & 3200 & 4312 \\
		7  & 1010 & 2143 & & 15 & 2110 & 3241 & & 23 & 3210 & 4321 \\
		\bottomrule
	\end{tabular}
\end{table}

\subsubsection{Fractal Block Characteristics of Tail States}
By analyzing the state evolution trajectory, we observe that the trailing two elements of the factorial array, $C[n-2 \dots n-1]$, exist in only two valid configurations:
\[
(0, 0) \quad \text{and} \quad (1, 0)
\]
These configurations manifest in a strict \textit{block-wise clustering} pattern across the state space. Physically, these transitions correspond directly to the pairwise transposition of elements in the underlying permutation, providing the boundary conditions for our SIMD-based acceleration.

\subsubsection{Empirical Observation: $n=4$ Lexicographical Mapping}
To visualize this fractal block-wise pattern, we track the state evolution for $n=4$. As illustrated in Table \ref{tab:factorial_4_full}, the periodicity of the trailing 3 elements reveals the clear boundaries suitable for hardware-level vectorization.

\clearpage

\begin{table}[htbp]
	\centering
	\caption{Full Lexicographical State Evolution and Periodic Block Structure for $n=4$}
	\label{tab:factorial_4_full}
	\begin{tabular}{ccccc}
		\toprule
		\textbf{Rank} & \textbf{Permutation} & \textbf{Factorial Array ($C$)} & \textbf{Suffix ($D_2, D_1, D_0$)} & \textbf{Block ID} \\
		\midrule
		0 & (0,1,2,3) & (0,0,0,0) & (0,0,0) & \\
		1 & (0,1,3,2) & (0,0,1,0) & (0,1,0) & \\
		2 & (0,2,1,3) & (0,1,0,0) & (1,0,0) & Block 0 \\
		3 & (0,2,3,1) & (0,1,1,0) & (1,1,0) & \\
		4 & (0,3,1,2) & (0,2,0,0) & (2,0,0) & \\
		5 & (0,3,2,1) & (0,2,1,0) & (2,1,0) & \\
		\addlinespace[10pt]
		6 & (1,0,2,3) & (1,0,0,0) & (0,0,0) & \\
		7 & (1,0,3,2) & (1,0,1,0) & (0,1,0) & \\
		8 & (1,2,0,3) & (1,1,0,0) & (1,0,0) & Block 1 \\
		9 & (1,2,3,0) & (1,1,1,0) & (1,1,0) & \\
		10 & (1,3,0,2) & (1,2,0,0) & (2,0,0) & \\
		11 & (1,3,2,0) & (1,2,1,0) & (2,1,0) & \\
		\addlinespace[10pt] 
		12 & (2,0,1,3) & (2,0,0,0) & (0,0,0) & \\
		13 & (2,0,3,1) & (2,0,1,0) & (0,1,0) & \\
		14 & (2,1,0,3) & (2,1,0,0) & (1,0,0) & Block 2 \\
		15 & (2,1,3,0) & (2,1,1,0) & (1,1,0) & \\
		16 & (2,3,0,1) & (2,2,0,0) & (2,0,0) & \\
		17 & (2,3,1,0) & (2,2,1,0) & (2,1,0) & \\
		\addlinespace[10pt]
		18 & (3,0,1,2) & (3,0,0,0) & (0,0,0) & \\
		19 & (3,0,2,1) & (3,0,1,0) & (0,1,0) & \\
		20 & (3,1,0,2) & (3,1,0,0) & (1,0,0) & Block 3 \\
		21 & (3,1,2,0) & (3,1,1,0) & (1,1,0) & \\
		22 & (3,2,0,1) & (3,2,0,0) & (2,0,0) & \\
		23 & (3,2,1,0) & (3,2,1,0) & (2,1,0) & \\
		\bottomrule
	\end{tabular}
\end{table}

\section{Mathematical Foundation: Fractal Block Structure} 

\subsection{Observation of Localized State Evolution}
Analyzing the state transitions in the factoradic system (e.g., from range $3000$ to $3210$), we observe a recurring structure in the trailing elements of the factorial array. Specifically, as the tail of the array evolves from $(0, 0, 0)$ to $(n-k-1, \dots, 2, 1, 0)$, the corresponding permutations manifest as complete lexicographical sequences of the remaining subset of elements. Regardless of the numerical values of the prefix, the elements within these localized blocks undergo a deterministic sequence of permutations, strictly adhering to lexicographical order. This $k!$ block structure repeats periodically, covering the entire state space of $n!$ permutations.

\begin{Observation}[Block Lexicographical Completeness]
	\label{prop:block_completeness}
	Given a $k$-element sub-block of the factorial array ranging from $(0, \dots, 0)$ to $(n-k-1, \dots, 1, 0)$, the set of permutations generated within this interval constitutes a complete lexicographical sequence of the involved $k$ elements, with a total cardinality of $k!$.
\end{Observation}

\begin{Observation}[Structural Isomorphism]
	For any $k!$ sub-block, the logic governing the state transition is invariant. By re-encoding the elements of any $k$-length prefix to a normalized sequence $\{1, 2, \dots, k\}$, the internal transition dynamics are isomorphic to the lexicographical permutation of the set $\{1, 2, \dots, k\}$.
\end{Observation}

\subsection{Design Implication: Universal SIMD Acceleration}
Building upon Observation 1 and Observation 2, we propose a generalized hardware acceleration logic. Since all $k!$ blocks share the same transformation logic, we can pre-define a universal SIMD permutation kernel. This kernel, once designed for $k$ elements, can be applied to all $n! / k!$ blocks, effectively offloading the entire permutation generation process to a constant-time hardware-accelerated primitive.

\section{LexCHA: Fractal-Based SIMD Acceleration}

Building upon the topological framework and fractal block properties established in Observation \ref{prop:block_completeness}, we now translate these theoretical foundations into the concrete hardware acceleration architecture of LexCHA.

\subsection{Precomputation of Transition Logic}
We pre-calculate the state-transition dynamics for $k=5$. Each transition is represented as a shuffle mask that transforms the current 128-bit SIMD register state into the subsequent lexicographical state.

\begin{algorithm}[h]
	\caption{Precomputation of the SIMD Transition LUT}
	\begin{algorithmic}[1]
		\Require $\text{TAIL\_DEPTH} = k$
		\State Initialize identity permutation $P \gets [0, 1, \dots, k-1]$
		\For{$\text{step} \gets 0$ to $k! - 1$}
		\State Let $P_{next} \gets \text{next\_permutation}(P)$
		\For{$j \gets 0$ to $k-1$}
		\State $\text{flat\_lut\_N5}[\text{step}][P_{next}[j]] \gets P[j]$
		\EndFor
		\State $P \gets P_{next}$
		\EndFor
	\end{algorithmic}
\end{algorithm}

\subsection{Inter-Block Transition and Amortized Analysis}

To bridge the transition between consecutive $k!$ sub-blocks, we employ a hybrid control flow. Upon reaching the terminal state of a $k!$ block, the hardware-level SIMD engine triggers an inter-block transition. This is achieved by executing a scalar \texttt{next\_permutation} on the global $n$-element sequence. 

\begin{Observation}[Amortized Efficiency]
	The generation of permutations relies on the $O(1)$ amortized cost of the standard next-permutation algorithm. Beyond this, our SIMD-accelerated framework further minimizes scalar overhead by processing $k!$ transitions in parallel. Consequently, the inter-block scalar transitions?which are inherent to the algorithm's control flow are effectively diluted, ensuring that the global throughput is primarily bounded by the peak SIMD compute capacity rather than sequential transition overhead.
\end{Observation}

This hybrid approach ensures that the performance bottleneck of the \texttt{next\_permutation} algorithm is minimized. The engine operates in a "macro-jump" fashion, where the scalar logic acts as a coarse-grained supervisor, while the hardware-hardened SIMD primitives provide fine-grained, branch-free execution for the overwhelming majority of permutation states.

\section{Overall Architecture Design of LexCHA}
\subsection{Design Philosophy: Fractal and Divide-and-Conquer}
\begin{figure}[htbp]
	\centering
	\includegraphics[width=0.8\textwidth]{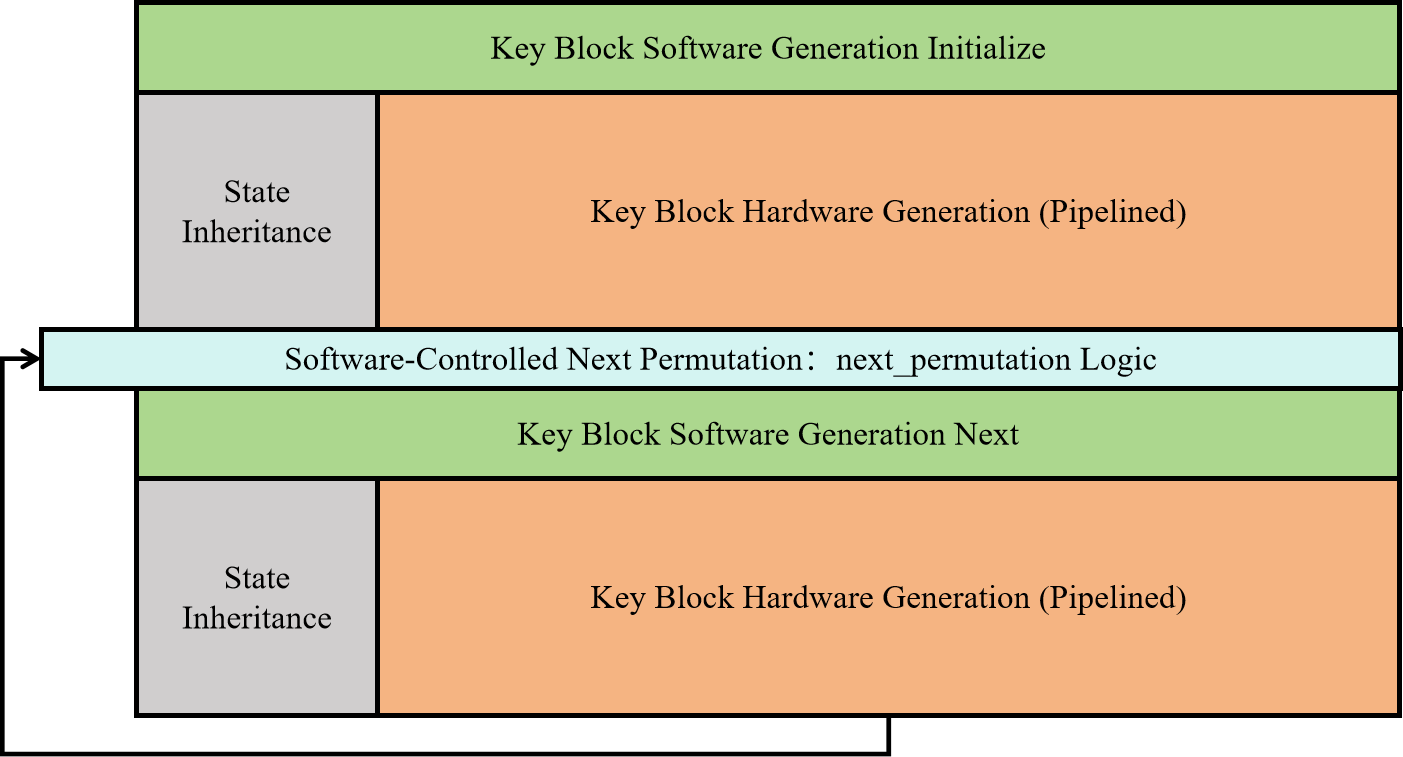}
	\caption{Co-design architecture of the LexCHA algorithm} \label{fig:lexcha_arch}
\end{figure}

LexCHA aims to reconstruct branch-heavy permutation tasks into branch-free dataflow computation. Lexicographical permutations exhibit inherent fractal properties. We separate the main sequence and data blocks for processing and adopt a hardware-software co-designed divide-and-conquer strategy, as illustrated in Fig. \ref{fig:lexcha_arch}. This design fully leverages the Shuffle instructions of modern CPUs. Instead of real-time dynamic suffix scanning, we employ precomputed permutation mappings to trade space overhead for deterministic execution, thereby maximizing hardware utilization.

\subsection{Logic Hardening}
Based on vector instruction primitives such as \texttt{PSHUFB} on the x86 architecture, permutation masks are directly loaded into vector registers. All operations are executed in parallel within the CPU arithmetic logic unit (ALU), achieving physical-level acceleration via hardware-accelerated execution.

We implement a Transition Buffer, whose core component is the Permutation Mask Table (PMT). With the fractal indexing mechanism, PMT only stores basic permutation subsets to reconstruct the evolution logic for permutations of any scale, realizing offline solidification of algorithm logic.

However, a critical trade-off emerges at $K=5$: the 120 permutations rely on only 10 ($= K(K-1)/2$) unique transformation masks. While compressing the PMT could save space, it necessitates runtime address indexing and branching. Our experiments show that the overhead of dynamic indexing, specifically index calculation latency and irregular memory access-outweighs the memory footprint benefits. Thus, we utilize a redundant storage strategy for $K=5$ to ensure a purely sequential execution flow. This approach eliminates runtime jumps, maximizing pipeline efficiency and throughput by maintaining a predictable dataflow.

\subsection{Algorithm Pseudocode}
Lexicographical permutations have fractal characteristics. We divide the computation into main sequence and sub-blocks, and adopt hardware-software co-designed divide-and-conquer. The proposed design makes full use of modern CPU Shuffle instructions.

\begin{algorithm}[H]
	\caption{LexCHA SIMD-Accelerated Permutation Engine}
	\label{alg:lexcha_accelerated}
	\begin{algorithmic}[1]
		\Procedure{LexCHA\_Engine}{$N$}
		\State $C \gets [0, 1, \dots, N-1]$ \Comment{Initialize sequence}
		\State $\mathit{buffer} \gets C[N - \mathit{TAIL\_DEPTH} \dots N-1]$
		\State $p\_reg \gets \text{LoadSIMD}(\mathit{buffer})$ \Comment{Load tail into XMM register}
		\State $\mathit{total\_count} \gets 1, \mathit{max\_perms} \gets N!$
		
		\While{$\mathit{total\_count} < \mathit{max\_perms}$}
		\For{$\mathit{step} \gets 0$ \textbf{to} $\mathit{FLAT\_STEPS} - 1$}
		\State $\mathit{mask} \gets \text{LoadSIMD}(\mathit{flat\_lut\_N5}[\mathit{step}])$
		\State $p\_reg \gets \text{SIMD\_Shuffle}(p\_reg, \mathit{mask})$ \Comment{Hardware-level transition}
		\EndFor
		\State $\mathit{total\_count} \gets \mathit{total\_count} + \mathit{FLAT\_STEPS}$
		
		\State $\mathit{buffer} \gets \text{StoreSIMD}(p\_reg)$ 
		\State $C[N - \mathit{TAIL\_DEPTH} \dots N-1] \gets \mathit{buffer}$ \Comment{Sync state to memory}
		
		\If{\Call{NextPermutation}{$C$}} \Comment{Coarse-grained scalar fallback}
		\State $\mathit{total\_count} \gets \mathit{total\_count} + 1$
		\State $\mathit{buffer} \gets C[N - \mathit{TAIL\_DEPTH} \dots N-1]$
		\State $p\_reg \gets \text{LoadSIMD}(\mathit{buffer})$ \Comment{Reload new state}
		\EndIf
		\EndWhile
		\State \Return $\mathit{total\_count}$
		\EndProcedure
	\end{algorithmic}
\end{algorithm}

\clearpage

\section{LexCHA (Lexicographical Co-designed Hardware Acceleration) Performance Benchmarks}
All benchmarks are restricted to lexicographical generation; our approach is specifically optimized for $n \ge 10$ on x86 platforms equipped with SSE/AVX extensions.

\subsection{Performance Comparison Test}
The Table \ref{tab:performance_comparison_1} and Table \ref{tab:performance_comparison_2} below present a performance comparison between the standard C++ library implementation (\texttt{std::next\_permutation}) and the SIMD-accelerated LexCHA engine on various CPU architectures.

\begin{itemize}
	\item Environment 1: Cloud VM (GitHub Actions / AMD EPYC)
	
	\item Compiler: \texttt{g++ -O3 -march=native -std=c++17}
	\item Optimization Target: Strict byte-level vectorization (\texttt{\_mm\_shuffle\_epi8})
\end{itemize}

\begin{table}[htbp]
	\centering
	\caption{Performance Comparison: Standard \texttt{std::next\_permutation} vs. LexCHA Cloud AMD}
	\label{tab:performance_comparison_1}
	\begin{tabular}{cccccc}
		\toprule
		\textbf{N} & \textbf{Std (s)} & \textbf{Acc (s)} & \textbf{Std (ns/perm)} & \textbf{Acc (ns/perm)} & \textbf{Speedup} \\
		\midrule
		10 & 0.031685 & 0.004391 & 8.73 & 1.21 & 7.22x \\
		11 & 0.357150 & 0.048634 & 8.95 & 1.22 & 7.34x \\
		12 & 4.486271 & 0.589570 & 9.37 & 1.23 & 7.61x \\
		13 & 60.799209 & 7.572245 & 9.76 & 1.22 & 8.03x \\
		\bottomrule
	\end{tabular}
\end{table}

\begin{itemize}
	\item Environment 2: Local Host (Intel Core Architecture)
	\item Compiler: g++ -O3 -march=native -std=c++17
	\item Dedicated Intel shuffle ports and optimal Store-to-Load Forwarding (STLF)
\end{itemize}

\begin{table}[htbp]
	\centering
	\caption{Performance Comparison: Standard \texttt{std::next\_permutation} vs. LexCHA Local Intel}
	\label{tab:performance_comparison_2}
	\begin{tabular}{lccccc}
		\toprule
		\textbf{N} & \textbf{Std (s)} & \textbf{Acc (s)} & \textbf{Std (ns/perm)} & \textbf{Acc (ns/perm)} & \textbf{Speedup} \\
		\midrule
		10 & 0.050085 & 0.003814 & 13.79 & 1.05 & 13.13$\times$ \\
		11 & 0.435035 & 0.038048 & 11.77 & 1.03 & 11.46$\times$ \\
		12 & 7.151700 & 0.429768 & 17.65 & 1.06 & 16.60$\times$ \\
		13 & 114.331200 & 6.793656 & 18.36 & 1.09 & 16.86$\times$ \\
		\bottomrule
	\end{tabular}
\end{table}

\subsection{Throughput Performance Analysis} 

Experimental results show that the runtime curve of LexCHA remains stable. Its per-permutation latency is nearly unaffected by the permutation length and maintains steady performance. 

In contrast, \texttt{std::next\_permutation} exhibits increasing per-permutation execution time as the value of $n$ increases. It is worth noting that while benchmark results indicate slight variations in optimization depth between AMD and Intel architectures, these differences are primarily attributable to hardware-specific branch prediction and instruction scheduling; they remain orthogonal to the fundamental algorithmic efficiency discussed in this study. 

The reference implementation, including the SIMD-optimized routines, is maintained in the \textit{Position-Pure-Algorithm} software repository. This repository provides the benchmark tools necessary to ensure the reproducibility of the performance data presented in Table \ref{tab:performance_comparison_1} and Table \ref{tab:performance_comparison_2}.

\section{Conclusion}
LexCHA thoroughly explores the topological rules behind the evolution of lexicographical permutations, and converts the traditional logic-intensive generation task into SIMD dataflow. This work establishes a hardware-software co-design paradigm of \textit{logic as data, execution as mapping}.



\end{document}